\documentclass{article}
\usepackage{spconf,amsmath,amssymb,graphicx}
\usepackage{xcolor}
\definecolor{mdown}{RGB}{33,102,172}  
\definecolor{mup}{RGB}{178,24,43}    

\usepackage{enumitem}
\setlist{nosep, leftmargin=14pt}

\title{ST-Dist: Distilling Spatial Transcriptomics Predictions into Interpretable Gene Axes for Distant Metastasis Prediction}
\name{Sandesh Pokhrel$^{1}$ \qquad Hamid Manoochehri$^{1}$ \qquad Beatrice S. Knudsen$^{2}$ \qquad Tolga Tasdizen$^{1}$}
\address{$^{1}$Department of Electrical and Computer Engineering and Scientific Computing and Imaging Institute,\\
University of Utah, Salt Lake City, UT, USA\\
$^{2}$Department of Pathology, University of Utah, Salt Lake City, UT, USA}
\begin{document}
\ninept
\maketitle
\begin{abstract}

Distant metastasis at diagnosis determines the treatment of prostate cancer, but predicting it from routine biopsy slides is challenging as the signs of metastasis are not clearly localized and are more subtle than in standard tumor detection and subtyping tasks. Multiple instance learning (MIL) on pathology foundation-model features has become the standard approach when labeled data are abundant, but such models remain hard to interpret. Distant metastasis is further challenging, because confirming the label requires clinical follow-up beyond the routine H\&E slides.  To this end, we propose ST-Dist, which distills the gene expression predicted by a spatial transcriptomics (ST) model into a linear head mapping each tile's foundation-model features onto named gene axes. The slide's projections along these interpretable gene axes are then used for distant metastasis prediction with a simple logistic regression. 

On 497 prostate needle biopsies (205 patients), ST-Dist reaches an AUROC of 91.37\%, above nine MIL baselines. It remains competitive on the public TCGA-COAD and TCGA-KIRC cohorts, and it has the highest mean AUROC over the three cohorts. The ablations on the prostate biopsies show that the gene axes outperform a linear probe on encoder features, dimension-matched principal components, random gene panels and the ST model's own predicted expression, and the gain over the probe holds on all 13 pathology encoder settings that we tested. ST-Dist thus offers a simple, accurate and interpretable route from H\&E to distant metastasis prediction.

\end{abstract}

\section{Introduction}
\label{sec:intro}
A patient who already has distant metastasis when prostate cancer is diagnosed (stage M1 rather than M0) needs systemic therapy, not local treatment alone. Knowing whether metastasis is present is therefore one of the most important questions at diagnosis, and one that the biopsy from the primary tumor alone does not yet answer. Unlike tumor detection or subtyping, which have abundant labels and a clearly localized signal, metastasis prediction relies on signs that are not confined to one region of the slide \cite{reactivestroma,fieldcancerization}. Additionally, confirming metastatic status requires further imaging and clinical follow-up, which makes labels expensive and cohorts small. 

Clinical risk groups and tissue-based genomic classifiers such as Decipher \cite{decipher}, the cell cycle progression score \cite{cellcycleprgressionscore} and the Genomic Prostate Score \cite{genomicprostatescore} help to assess the aggressiveness of the cancer but cannot establish a clear metastatic status. These classifiers additionally need dedicated molecular assays, which are not always available. In contrast, the routine hematoxylin and eosin (H\&E) slide exists for every patient, but whether this slide can predict M0/M1 status at diagnosis is largely unexplored.

Recently, deep learning has shown the prognostic value of biopsy histology by combining slide images with clinical variables to predict long-term distant metastasis in phase III trial cohorts \cite{prostatecancertherapypersonalization}. Standard approaches like multiple instance learning (MIL) on pathology foundation-model features \cite{abmil,rrtmil,dtmf-mil} learn attention weights that pick out the most informative tiles, which suits localized signals. With few labels, however, simple pooling without attention is often as good as complex MIL \cite{simlp,nonattnmil}. Further, the explanations that MIL offers, mostly attention heatmaps, are not always accepted as interpretable \cite{labelfreemil}. 

Gene expression offers a different language for the same slide. Clinicians already reason about metastatic risk in terms of genes, and with models now capable of predicting expression from H\&E, predicted expression can be leveraged for downstream tasks. Paired spatial transcriptomics (ST) resources such as HEST-1k \cite{hest1k} have enabled the development of foundation models such as STPath \cite{stpath}, which predicts tens of thousands of genes for every tile of a WSI. Similarly, virtual proteins predicted from H\&E have been shown to improve prognostic prediction in lung cancer~\cite{hex}, further showing that predicted expression is useful for prognosis.

In this work, we propose genes as a concept bottleneck \cite{conceptbottleneck}. Unlike a recent concept-based MIL method~\cite{labelfreemil}, which predicts its concepts with a vision-language model, we distill the concepts (genes) from an ST foundation model, STPath. A linear head on top of a frozen pathology encoder learns to reproduce the expression that the teacher, STPath, predicts for a small, pre-specified panel of metastasis related genes. A logistic regression trained on the outcome labels then predicts metastasis from the slide-averaged gene values. The head can sit on any pathology encoder, independent of the teacher's backbone. We call this model ST-Dist. Every input of downstream classifiers using ST-Dist is thus a named gene rather than an unnamed image feature.

On 497 prostate needle biopsies, ST-Dist outperforms nine MIL methods, and we further evaluate it on two public cohorts of The Cancer Genome Atlas (TCGA), TCGA-COAD and TCGA-KIRC. Our contributions can be summarized as: 

\begin{itemize}
    \item We introduce ST-Dist, an encoder-agnostic way to read pathology features through named gene axes, distilled from an ST model without outcome labels.
    \item We show on three metastasis cohorts that this simple model matches or exceeds nine MIL methods, and that its gain over a linear probe on encoder features holds on 13 different encoder on prostate biopsies.
    \item We trace the gain to the choice of genes rather than to the reduction in dimension, and validate STPath's predictions of two of the genes with the largest M1--M0 differences, MSMB and AZGP1, on measured ST.
\end{itemize}

\section{Method}
\label{sec:method}

\begin{figure}[t]
  \centering
  \includegraphics[width=\linewidth]{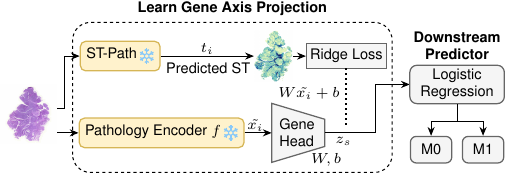}
  \caption{Overview of ST-Dist. \textit{Learn Gene Axis Projection:} without outcome labels, the gene head $(W, \mathbf{b})$ is fit so that its tile values $W\tilde{\mathbf{x}}_i + \mathbf{b}$ match STPath's predicted expression $\mathbf{t}_i$ (ridge loss, Eq.~\ref{eq:ridge}). \textit{Downstream Predictor:} with the gene head fixed, a logistic regression predicts M0 or M1 from the standardized, slide-averaged gene values $\mathbf{z}_s$. \textit{Snowflake: frozen. STPath is not used at inference.}}
  \label{fig:overview}
\end{figure}

\textbf{Problem setting.} A slide $s$ is represented by its $N_s$ tissue tiles, with features $\mathbf{x}_i = f(\text{tile}_i) \in \mathbb{R}^d$ from a frozen pathology encoder $f$. The task is to predict the slide label $y_s \in \{0, 1\}$ (M0 or M1) from a small labeled cohort. ST-Dist addresses it as shown in Fig.~\ref{fig:overview}.

\noindent \textbf{Tile encoder.} Our primary pathology encoder ($f$) is Prov-Giga\-Path~\cite{provgigapath}, a vision transformer pretrained by self-supervised learning on 1.3~billion tiles ($d = 1536$). Any other pathology encoder can serve as $f$. The teacher, STPath, always receives GigaPath features, so a different encoder changes only the input of the gene head.

\noindent \textbf{Spatial transcriptomics teacher.} STPath is a generative foundation model that predicts the expression of 38{,}984 genes from histology. It is a geometry-aware transformer, pretrained by masked gene-expression prediction on about 1{,}000 public slides with paired ST profiles from 17 organs. Each tile is one token, which combines the tile's GigaPath feature, an expression input, an organ token and a sequencing-technology token; tile positions enter through the geometry-aware attention. At inference, all expression inputs are masked, so the prediction rests on histology alone. We use the organ token of each cohort (Prostate, Kidney or Bowel) and no technology token. For tile $i$, STPath produces a 512-dimensional latent $\mathbf{u}_i$ (after layer normalization). Each gene $g$ is a fixed linear readout of it, $t_{ig} = \mathbf{a}_g^\top \mathbf{u}_i + a_{g0}$. For a panel of $G$ genes, we collect these values in $\mathbf{t}_i \in \mathbb{R}^G$.

\noindent \textbf{Choosing the gene panel.}  
The gene panel is a predefined set of genes to predict from the teacher. It is fixed before any outcome is seen and is derived from the union of published gene sets for the endpoint: signatures of metastasis and aggressiveness, lineage genes and cancer drivers. Genes outside the teacher's vocabulary are removed. Our selection of panels is described in Sec.~\ref{sec:experiments}.

\noindent \textbf{Distilling the gene head.} The student is a linear gene head, $W \in \mathbb{R}^{G \times d}$ and $\mathbf{b} \in \mathbb{R}^G$, that maps a tile feature to the teacher's values. Let $\tilde{\mathbf{x}}_i$ denote the standardized feature of tile $i$. Row $\mathbf{w}_g$ of $W$ is the axis of gene $g$ in feature space, and the tile's value for gene $g$ is its coordinate along this axis, $\mathbf{w}_g^\top \tilde{\mathbf{x}}_i + b_g$. Over the set $\mathcal{S}$ of training slides, the head minimizes the slide-weighted ridge objective
\begin{equation}
\frac{1}{|\mathcal{S}|} \sum_{s \in \mathcal{S}} \frac{1}{N_s} \sum_{i \in s} \big\| \mathbf{t}_i - W \tilde{\mathbf{x}}_i - \mathbf{b} \big\|_2^2 + \lambda \| W \|_F^2,
\label{eq:ridge}
\end{equation}
so that every WSI has the same total weight, whatever its size. The regularization parameter $\lambda$ is chosen from ten values between $10^{-4}$ and $10^{2}$, by the mean per-gene $R^2$ between head and teacher on the validation slides. No outcome label enters this step, and the head is refit in every fold. 

\noindent \textbf{Slide gene values.} For slide $s$ with mean standardized tile feature $\bar{\mathbf{x}}_s = \frac{1}{N_s} \sum_{i \in s} \tilde{\mathbf{x}}_i$, the slide's gene values $\hat{\mathbf{e}}_s \in \mathbb{R}^G$ are the average of the tile-level gene values,
\begin{equation}
\hat{\mathbf{e}}_s = W \bar{\mathbf{x}}_s + \mathbf{b} = \frac{1}{N_s} \sum_{i \in s} \big( W \tilde{\mathbf{x}}_i + \mathbf{b} \big).
\label{eq:axes}
\end{equation}
Each component of $\hat{\mathbf{e}}_s$ is standardized with the mean and standard deviation over the training slides, which gives $\mathbf{z}_s \in \mathbb{R}^G$.

\noindent \textbf{Metastasis classifier.} The classifier is an $\ell_2$-regularized logistic regression on $\mathbf{z}_s$, with inverse regularization strength $C = 1$. This value was fixed in advance and never tuned, so the validation slides serve only to choose the ridge penalty. At inference, a slide needs only the encoder, the gene head and the classifier. The teacher is used only to train the head.

\noindent \textbf{Why concept bottleneck.} Because $\mathbf{z}_s$ is an affine function of $\bar{\mathbf{x}}_s$, ST-Dist is a linear probe restricted to the subspace spanned by the gene axes, of dimension at most $G$. The genes thus act as a prior on which visual features may matter. 

\section{Experiments}
\label{sec:experiments}

\textbf{Cohorts.} Our main cohort consists of 497 pretreatment prostate needle biopsies from 205 patients (98 with M1), labeled M0 or M1 for distant metastasis from clinical reports and follow-up. The main cohort data come from a large US hospital network. To test our approach on other cancers, we use two TCGA resection cohorts \cite{tcga} with the same endpoint: kidney clear-cell carcinoma (TCGA-KIRC, 509 slides from 505 patients, 76 with M1) and colon adenocarcinoma (TCGA-COAD, 387 slides from 380 patients, 60 with M1). The TCGA labels use the pathologic M stage (the clinical one when it is missing). Patients with MX or no M stage are excluded. Each cohort is split into three patient-level folds, each with training, validation and test sets.

\noindent \textbf{Gene panels.} Each cohort has its own panel. The prostate panel has 117 genes. Of these, 100 come from published signatures of metastasis and aggressiveness \cite{metastaticlethal,decipher,cellcycleprgressionscore,genomicprostatescore}. Ten are prostate lineage genes, such as NKX3-1 and MSMB, and seven are genes often altered in metastases. The kidney (98 genes) and colon (113 genes) panels come from five and seven published sources. The panel files, with the source of every gene, will be released with the code.

\noindent \textbf{A second vocabulary.} To test whether the result depends on our hand-built panel, we also use a vocabulary from a single standard source, fixed by name before its result was computed: the 1{,}236 genes of 10 Molecular Signatures Database (MSigDB) \textit{Hallmark pathways} linked to metastasis~\cite{msigdbhallmark}. These are epithelial--mesenchymal transition, TGF-$\beta$ signaling, apical junction, hypoxia, angiogenesis, inflammatory response, IL6/JAK/STAT3 signaling, TNF$\alpha$ signaling via NF-$\kappa$B, E2F targets and the G2M checkpoint. 


\noindent \textbf{Encoders and baselines.} Our main experiments use GigaPath features of 62.5~$\mu$m tiles, and robustness is tested on 13 settings of 9 pathology encoders (Sec.~\ref{sec:results}). The other settings use 125~$\mu$m tiles. For a 125~$\mu$m tile, the teacher's latent is averaged over the four 62.5~$\mu$m tiles that it contains, and the gene values follow by the same linear readout. We compare ST-Dist with nine MIL methods. Seven use attention: ABMIL~\cite{abmil}, CLAM~\cite{clam}, DSMIL~\cite{dsmil}, TransMIL~\cite{transmil}, DTFD-MIL~\cite{dtfdmil}, ILRA~\cite{ilra} and RRT-MIL~\cite{rrtmil}. Two use simple pooling techniques: SiMLP~\cite{simlp} combines mean-pooling with MLP and Maxsoft~\cite{nonattnmil} takes the highest tile score. All MIL methods share one training recipe: AdamW with a learning rate of $2 \times 10^{-4}$ and a weight decay of $10^{-4}$, dropout of 0.25 and 20 epochs, keeping the epoch with the highest validation area under the ROC curve (AUROC), with five random seeds per fold. Our main ablation removes the gene head: a mean-pooling linear probe applies the same logistic regression to the slide mean of all encoder features. Random gene panels test whether any genes that the teacher predicts would work as well as the chosen ones. For the curated panel and the Hallmark vocabulary alike, each of 200 random counterparts replaces every gene by one of its 100 nearest genes outside the vocabulary in predicted variance, so that the random genes vary as much across tiles as the original predicted genes.

\noindent \textbf{Evaluation.} We report the slide-level AUROC as the mean over the three test folds, with its standard deviation (SD) over runs (three folds for logistic models, 15 fold-and-seed runs for MIL). On the main cohort we also report average precision (AP) and balanced accuracy at a 0.5 threshold. All metrics are given in percent. 
Differences between models are tested with a paired bootstrap over patients (2{,}000 resamples), in which predictions are averaged over seeds and over each patient's slides, and pooled over the three test folds. We call a difference significant when its 95\% bootstrap confidence interval excludes zero.

\section{Results}
\label{sec:results}

\subsection{Accuracy, ablations and robustness}

\noindent \textbf{Performance on the prostate biopsies.} 
Table~\ref{tab:va} compares ST-Dist with nine MIL methods on the main cohort. ST-Dist has the highest AUROC (91.37\%), average precision (92.29\%) and balanced accuracy (82.12\%), and its AUROC is 2.2 points above the best MIL method (ABMIL, 89.13\%). Among the MIL methods, simple mean pooling (SiMLP, 88.54\%) comes within 0.6 points of the best attention-based method, in line with earlier findings for small cohorts~\cite{simlp,nonattnmil}.
\begin{table}[t]
\centering
\caption{Prostate biopsies (the main cohort), in percent: mean $\pm$ SD over runs. \textbf{Bold}: best; \underline{underline}: second.} 
\label{tab:va}
\resizebox{\linewidth}{!}{%
\begin{tabular}{lccc}
\hline
Method & AUROC & AP & Balanced acc. \\
\hline
\textbf{ST-Dist (ours)} & \textbf{91.37$\pm$2.19} & \textbf{92.29$\pm$1.60} & \textbf{82.12$\pm$2.15} \\
ILRA & 87.63$\pm$2.89 & 88.56$\pm$3.29 & 78.98$\pm$4.40 \\
ABMIL & \underline{89.13$\pm$2.28} & \underline{90.16$\pm$2.90} & 80.28$\pm$4.44 \\
RRT-MIL & 89.08$\pm$2.37 & 89.88$\pm$2.78 & 76.86$\pm$5.82 \\
CLAM & 88.34$\pm$2.47 & 89.34$\pm$2.83 & 79.53$\pm$4.30 \\
DSMIL & 86.98$\pm$3.31 & 87.99$\pm$3.59 & 77.82$\pm$4.83 \\
TransMIL & 83.94$\pm$3.84 & 84.55$\pm$4.38 & 71.35$\pm$7.47 \\
DTFD-MIL & 72.63$\pm$8.59 & 72.39$\pm$8.94 & 66.52$\pm$8.10 \\
SiMLP & 88.54$\pm$1.57 & 89.78$\pm$2.40 & \underline{80.56$\pm$2.42} \\
Maxsoft & 84.23$\pm$2.25 & 85.07$\pm$2.84 & 74.36$\pm$2.95 \\
\hline
\end{tabular}}
\end{table}

\noindent \textbf{Results on the TCGA cohorts.} We refit the same recipe on each TCGA cohort, with the cohort's own gene panel (Table~\ref{tab:tcga}). These resection cohorts are harder than the biopsies: M1 patients are rarer (15--16\%), and every method scores lower. On TCGA-COAD, ST-Dist (67.71\%) outperforms all nine MIL methods, by 2.9 points over the best one (Maxsoft), significantly so against RRT-MIL, TransMIL and DTFD-MIL. On TCGA-KIRC, it ranks second (81.19\%), 0.2 points behind ILRA. 
Without the gene head, the linear probe reaches 78.05\% on TCGA-KIRC and 69.16\% on TCGA-COAD. Overall, ST-Dist has the highest mean AUROC (80.09\%, against 78.68\% for the probe and 77.36\% for the best MIL method, CLAM).

\begin{table}[t]
\centering
\caption{TCGA cohorts (resections; 15--16\% M1 patients): AUROC in percent, mean $\pm$ SD as in Table~\ref{tab:va}. The mean is over the three cohorts, with the prostate biopsies from Table~\ref{tab:va}. \textbf{Bold}: best; \underline{underline}: second.}
\label{tab:tcga}
\setlength{\tabcolsep}{4pt}
\begin{tabular}{@{}lccc@{}}
\hline
Method & TCGA-KIRC & TCGA-COAD & Mean (3 cohorts) \\
\hline
\textbf{ST-Dist (ours)} & \underline{81.19$\pm$0.52} & \textbf{67.71$\pm$1.52} & \textbf{80.09} \\
ILRA & \textbf{81.38$\pm$4.80} & 62.63$\pm$5.04 & 77.21 \\
ABMIL & 79.76$\pm$5.04 & 61.35$\pm$6.18 & 76.75 \\
RRT-MIL & 80.62$\pm$4.56 & 61.39$\pm$4.62 & 77.03 \\
CLAM & 80.10$\pm$4.75 & 63.64$\pm$4.99 & \underline{77.36} \\
DSMIL & 75.79$\pm$4.96 & 61.85$\pm$7.30 & 74.87 \\
TransMIL & 72.06$\pm$6.67 & 58.81$\pm$6.97 & 71.60 \\
DTFD-MIL & 65.66$\pm$7.50 & 51.66$\pm$7.05 & 63.32 \\
SiMLP & 78.71$\pm$2.97 & 63.62$\pm$6.00 & 76.96 \\
Maxsoft & 78.67$\pm$2.97 & \underline{64.77$\pm$5.59} & 75.89 \\
\hline
\end{tabular}
\end{table}

\noindent \textbf{Source of the gain.} Table~\ref{tab:ablations} tests three explanations of the gain. (i)~\textit{Compression}: if ST-Dist only compressed the 1,536 features into 117, any 117-dimensional projection would work as well. It does not. The 117 principal components fall below the linear probe (86.78\% against 88.82\%), and random gene panels of the same size reach only 89.49\%, with the curated panel above 97\% of them. (ii)~\textit{The teacher's representation}: distilling the full 512-dimensional STPath latent instead of the genes gives almost no gain over the probe (89.03\%), so the teacher helps mainly through the chosen genes. (iii)~\textit{Restriction}: adding the raw features back to the gene values (88.96\%) returns the AUROC to the level of the probe, because the classifier can again use every visual feature direction. The benefit therefore comes from the choice of genes, not from the reduction in dimension. Two further rows test design choices. Learning tile weights with attention on top of ST-Dist's classifier gives a similar AUROC (91.53\%), and the learned weights stay close to equal (on average, they spread over 91\% of the tiles), so the plain slide average suffices. A nonlinear gene head (an MLP with two hidden layers of 1,024 units) reproduces the teacher more closely than a linear layer trained in the same label-free way (validation $R^2$ 0.729 against 0.691), but its AUROC on the prostate biopsies is similar, so we keep the simpler linear head. 

\begin{table}[t]
\centering
\caption{Ablations on the prostate biopsies (AUROC in percent, mean $\pm$ SD as in Table~\ref{tab:va}; for random gene panels, SD over the 200 panels). \textbf{Bold}: best; \underline{underline}: second.}
\label{tab:ablations}
\begin{tabular}{lc}
\hline
Model & AUROC \\
\hline
\textbf{ST-Dist} & 91.37$\pm$2.19 \\
Nonlinear gene head (MLP) & \textbf{91.68$\pm$2.78} \\
Attention-weighted ST-Dist & \underline{91.53$\pm$1.45} \\
Hallmark vocabulary & 90.81$\pm$0.35 \\
STPath expression & 89.60$\pm$2.73 \\
Random gene panels & 89.49$\pm$1.02 \\
ST-Dist on the full latent (512) & 89.03$\pm$0.35 \\
ST-Dist + raw features & 88.96$\pm$1.57 \\
Linear probe (no gene head) & 88.82$\pm$1.80 \\
PCA-117 & 86.78$\pm$0.81 \\
\hline
\end{tabular}
\end{table}

\noindent \textbf{Robustness to the choice of vocabulary.} 
The gain does not depend on our hand-built panel. Replacing it with the genes of the 10 metastasis-related Hallmark pathways (1,236 genes, of which only 41 are in the 117-panel) yields an AUROC of 90.81 $\pm$ 0.35\%, above the linear probe (88.82\%) and above 97\% of its 200 variance-matched random counterparts.

\noindent \textbf{The student outperforms its teacher.} ST-Dist retains only the part of STPath's prediction that a tile's own features explain linearly. In all three cohorts, this part predicts metastasis better than STPath's own prediction of the same genes (STPath expression): 91.37\% against 89.60\% on the prostate biopsies, 81.19\% against 78.90\% on TCGA-KIRC and 67.71\% against 62.69\% on TCGA-COAD, where the difference is significant.

\noindent \textbf{Robustness across encoders.} ST-Dist outperforms the linear probe on all 13 encoder settings, by 0.10 to 4.69 points (Table~\ref{tab:encoders}). 
Encoders that reproduce the teacher more closely tend to give better ST-Dist models (Spearman 0.72 between the validation $R^2$ and the ST-Dist AUROC). The curated panel ranks at or above the 90th percentile of variance-matched random gene panels on 9 of the 13 settings.

\begin{table}[t]
\centering
\caption{ST-Dist on 13 settings of 9 pathology encoders (prostate biopsies, AUROC in percent, mean $\pm$ SD over 3 folds), sorted by the linear probe. $R^2$: mean validation $R^2$ of the gene head against the teacher. CLS + mean: the class token concatenated with the mean patch token. Encoders: GigaPath~\cite{provgigapath}, UNI and UNI2-h~\cite{Uni}, Virchow~\cite{virchow}, Virchow2~\cite{virchow2}, Phikon~\cite{phikon}, Phikon-v2~\cite{phikonv2}, Hibou-B~\cite{hibou} and MoCo v3~\cite{mocov3}.}
\label{tab:encoders}
\setlength{\tabcolsep}{4pt}
\begin{tabular}{@{}lccc@{}}
\hline
Encoder & $R^2$ & Linear probe & ST-Dist \\
\hline
MoCo v3 ResNet-50 (SRA)$^\dagger$ & 0.508 & 82.20$\pm$2.28 & 86.89$\pm$4.73 \\
MoCo v3 ResNet-50$^\dagger$ & 0.537 & 83.73$\pm$2.33 & 88.38$\pm$3.50 \\
Virchow2 (CLS + mean) & 0.636 & 86.95$\pm$3.20 & 91.48$\pm$2.24 \\
Hibou-B & 0.594 & 87.82$\pm$2.13 & 89.85$\pm$2.45 \\
Virchow2 (CLS) & 0.622 & 88.67$\pm$2.33 & 91.05$\pm$2.32 \\
\textbf{GigaPath (62.5 $\mu$m, main)} & 0.694 & 88.82$\pm$1.80 & 91.37$\pm$2.19 \\
GigaPath (125 $\mu$m) & 0.633 & 89.66$\pm$0.38 & 91.06$\pm$2.62 \\
Phikon & 0.582 & 89.68$\pm$1.80 & 90.48$\pm$1.72 \\
Virchow (CLS + mean) & 0.636 & 89.80$\pm$0.89 & 91.23$\pm$1.99 \\
Virchow (CLS) & 0.611 & 89.85$\pm$2.03 & 90.76$\pm$2.35 \\
UNI & 0.599 & 89.97$\pm$1.33 & 91.95$\pm$2.03 \\
Phikon-v2 & 0.603 & 90.58$\pm$1.78 & 90.68$\pm$1.43 \\
UNI2-h & 0.614 & 91.62$\pm$0.71 & 92.06$\pm$1.69 \\
\hline
\end{tabular}
\\[2pt]
\parbox{\linewidth}{$^\dagger$ Two in-house runs, self-supervised on about 60\% of TCGA~\cite{tcga} slides from all organs. Both use the official MoCo v3 augmentations; SRA (stain reconstruction augmentation~\cite{sra}) adds one that randomizes the amounts of the H and E stains separately.}
\end{table}

\subsection{Which genes separate M1 from M0}

\begin{figure}[!t]
  \centering
  \includegraphics[width=\linewidth]{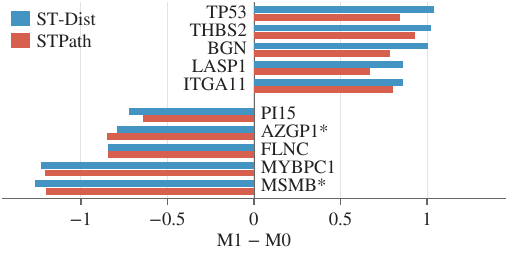}
  \caption{The largest M1--M0 differences in ST-Dist and STPath (Hedges' $g$; 98 M1, 107 M0 patients), all significant in both (Mann--Whitney test, $p < 10^{-3}$, corrected for 117 genes).\textit{ $^*$ marks STPath's prediction validated on measured ST.}}
  \label{fig:genes}
\end{figure}

\noindent \textbf{Genes with the largest differences.} ST-Dist outperforms its teacher (Table~\ref{tab:ablations}). To check that its gene values still carry the teacher's biology, Fig.~\ref{fig:genes} compares the ten genes with the largest M1--M0 differences in ST-Dist with the same genes in STPath, measured by Hedges' $g$ (the difference in patient means, divided by the pooled SD). ST-Dist's differences closely follow STPath's (Pearson r = 0.99 over all 117 genes) and are larger for 8 of the 10 genes. The student thus keeps the teacher's gene-level signal and separates M1 from M0 slightly more strongly.

\noindent \textbf{Grounding in measured expression.} On eight held-out prostate Visium slides from HEST-Bench~\cite{hest1k}, which STPath did not see during training, STPath predicts 12 of the 115 panel genes that these slides measure with a mean correlation of at least 0.3 (0.32 to 0.52) against measured expression, computed per gene within each slide. These are mostly luminal and androgen-regulated genes (MSMB, AZGP1, SLC45A3, TMPRSS2, KLK2, KLK3, HOXB13, NKX3-1, SPDEF, ANO7, TPM2 and FOXA1). They include two of the genes with the largest M1--M0 differences in both models, MSMB and AZGP1 (Fig.~\ref{fig:genes}), whose loss has been associated with aggressive prostate cancer~\cite{azgp1,msmb}. Both genes are lower in M1 patients, which agrees with these reports.

\noindent \textbf{Recovery of known biology.} Used alone in place of the panel, the E2F targets pathway (one of the 10 Hallmark pathways) reaches an AUROC of 91.43 $\pm$ 1.07\%. This is at the 96th percentile of 200 variance-matched random gene sets of the same size (nominal $p = 0.05$).
E2F targets form the proliferation program that the cell cycle progression score~\cite{cellcycleprgressionscore} also measures, a score already used clinically to assess the aggressiveness of prostate cancer.


\section{Discussion and Conclusion}
\label{sec:discussion}

ST-Dist shows that distant metastasis, a label-scarce target, is well served by a simple model that reads pathology features through a small set of named genes rather than unnamed image features. The genes do more than make the model interpretable, they act as a prior that selects which visual directions are worth learning from few labels. A task-relevant vocabulary, whether our curated panel or the metastasis-related Hallmark pathways, keeps most of this benefit. In summary, distilling an ST foundation model into named gene axes turns any pathology encoder into a simple, interpretable and competitive predictor of distant metastasis from routine H\&E slides.

\section{Compliance with Ethical Standards}
\label{sec:ethics}
As an analysis of de-identified images, the study is not human subjects research and was exempt from IRB review. The TCGA~\cite{tcga} and HEST-1k~\cite{hest1k} data are publicly available, and their use did not require separate ethical approval. 

\section{Acknowledgments}
\label{sec:ack}
This work was supported by U.S. Department of Defense (DoD) HT94252410186. We declare no conflicts of interest. We used Claude (Anthropic) to draft code and edit language. ChatGPT Pro(Open AI) was used to curate gene panels.

\bibliographystyle{IEEEbib}
\bibliography{strings,refs}

@article{decipher,
  author = {Erho, Nicholas and others},
  title = {Discovery and validation of a prostate cancer genomic classifier that predicts early metastasis following radical prostatectomy},
  journal = {PLoS One},
  volume = {8},
  number = {6},
  pages = {e66855},
  month = jun,
  year = {2013}
}

@article{cellcycleprgressionscore,
  author = {Cuzick, Jack and others},
  title = {Prognostic value of an {RNA} expression signature derived from cell cycle proliferation genes in patients with prostate cancer: a retrospective study},
  journal = {Lancet Oncol.},
  volume = {12},
  number = {3},
  pages = {245--255},
  month = mar,
  year = {2011}
}

@article{genomicprostatescore,
  author = {Klein, Eric A and others},
  title = {A 17-gene assay to predict prostate cancer aggressiveness in the context of {Gleason} grade heterogeneity, tumor multifocality, and biopsy undersampling},
  journal = {Eur. Urol.},
  volume = {66},
  number = {3},
  pages = {550--560},
  month = may,
  year = {2014}
}

@article{prostatecancertherapypersonalization,
  author = {Esteva, Andre and others},
  title = {Prostate cancer therapy personalization via multi-modal deep learning on randomized phase {III} clinical trials},
  journal = {npj Digit. Med.},
  volume = {5},
  number = {1},
  pages = {71},
  month = jun,
  year = {2022}
}

@inproceedings{abmil,
  author = {Ilse, Maximilian and Tomczak, Jakub and Welling, Max},
  title = {Attention-based Deep Multiple Instance Learning},
  booktitle = {Proc. ICML},
  pages = {2127--2136},
  year = {2018}
}

@article{simlp,
  author = {Li, Jiawen and others},
  title = {Can We Simplify Slide-level Fine-tuning of Pathology Foundation Models?},
  journal = {arXiv:2502.20823},
  year = {2025}
}

@article{sra,
  author = {Manoochehri, Hamid and Zhang, Bodong and Knudsen, Beatrice S. and Tasdizen, Tolga},
  title = {{SRA}: A Novel Method to Improve Feature Embedding in Self-supervised Learning for Histopathological Images},
  journal = {arXiv:2410.17514},
  year = {2024}
}

@article{dtmf-mil,
  author = {Pokhrel, Sandesh and Manoochehri, Hamid and Knudsen, Beatrice S. and Tasdizen, Tolga},
  title = {Predicting Metastatic Risk from Primary Cancer Tissue Architecture via Distance-Aware Spatial Modeling},
  journal = {arXiv:2606.28676},
  year = {2026}
}

@inproceedings{nonattnmil,
  author = {Jafarinia, Hossein and others},
  title = {Navigating the {MIL} Trade-Off: Flexible Pooling for Whole Slide Image Classification},
  booktitle = {Proc. NeurIPS},
  pages = {96876--96921},
  year = {2025}
}

@article{labelfreemil,
  author = {Sun, Susu and others},
  title = {Label-free Concept Based Multiple Instance Learning for Gigapixel Histopathology},
  journal = {arXiv:2501.02922},
  year = {2025}
}

@inproceedings{hest1k,
  author = {Jaume, Guillaume and others},
  title = {{HEST-1k}: A Dataset For Spatial Transcriptomics and Histology Image Analysis},
  booktitle = {Proc. NeurIPS},
  pages = {53798--53833},
  year = {2024}
}

@article{stpath,
  author = {Huang, Tinglin and Liu, Tianyu and Babadi, Mehrtash and Ying, Rex and Jin, Wengong},
  title = {{STPath}: a generative foundation model for integrating spatial transcriptomics and whole-slide images},
  journal = {npj Digit. Med.},
  volume = {8},
  number = {1},
  pages = {659},
  month = nov,
  year = {2025}
}

@inproceedings{conceptbottleneck,
  author = {Koh, Pang Wei and others},
  title = {Concept Bottleneck Models},
  booktitle = {Proc. ICML},
  pages = {5338--5348},
  year = {2020}
}

@article{clam,
  author = {Lu, Ming Y. and Williamson, Drew F. K. and Chen, Tiffany Y. and Chen, Richard J. and Barbieri, Matteo and Mahmood, Faisal},
  title = {Data-efficient and weakly supervised computational pathology on whole-slide images},
  journal = {Nat. Biomed. Eng.},
  volume = {5},
  number = {6},
  pages = {555--570},
  year = {2021}
}

@inproceedings{transmil,
  author = {Shao, Zhuchen and others},
  title = {{TransMIL}: Transformer based Correlated Multiple Instance Learning for Whole Slide Image Classification},
  booktitle = {Proc. NeurIPS},
  pages = {2136--2147},
  year = {2021}
}

@inproceedings{dsmil,
  author = {Li, Bin and Li, Yin and Eliceiri, Kevin W.},
  title = {Dual-stream Multiple Instance Learning Network for Whole Slide Image Classification with Self-supervised Contrastive Learning},
  booktitle = {Proc. CVPR},
  pages = {14313--14323},
  year = {2021}
}

@inproceedings{dtfdmil,
  author = {Zhang, Hongrun and others},
  title = {{DTFD-MIL}: Double-Tier Feature Distillation Multiple Instance Learning for Histopathology Whole Slide Image Classification},
  booktitle = {Proc. CVPR},
  pages = {18780--18790},
  year = {2022}
}

@inproceedings{ilra,
  author = {Xiang, Jinxi and Zhang, Jun},
  title = {Exploring Low-Rank Property in Multiple Instance Learning for Whole Slide Image Classification},
  booktitle = {Proc. ICLR},
  year = {2023}
}

@inproceedings{rrtmil,
  author = {Tang, Wenhao and Zhou, Fengtao and Huang, Sheng and Zhu, Xiang and Zhang, Yi and Liu, Bo},
  title = {Feature Re-Embedding: Towards Foundation Model-Level Performance in Computational Pathology},
  booktitle = {Proc. CVPR},
  pages = {11343--11352},
  year = {2024}
}

@article{Uni,
  author = {Chen, Richard J. and others},
  title = {Towards a general-purpose foundation model for computational pathology},
  journal = {Nat. Med.},
  volume = {30},
  number = {3},
  pages = {850--862},
  year = {2024}
}

@article{provgigapath,
  author = {Xu, Hanwen and others},
  title = {A whole-slide foundation model for digital pathology from real-world data},
  journal = {Nature},
  volume = {630},
  number = {8015},
  pages = {181--188},
  year = {2024}
}

@article{hex,
  author = {Li, Zhe and others},
  title = {{AI-enabled} virtual spatial proteomics from histopathology for interpretable biomarker discovery in lung cancer},
  journal = {Nat. Med.},
  volume = {32},
  number = {1},
  pages = {231--244},
  year = {2026}
}

@article{metastaticlethal,
  author = {Rubicz, Rohina and others},
  title = {Gene expression panel predicts metastatic-lethal prostate cancer outcomes in men diagnosed with clinically localized prostate cancer},
  journal = {Mol. Oncol.},
  volume = {11},
  number = {2},
  pages = {140--150},
  year = {2017}
}

@article{msigdbhallmark,
  author = {Liberzon, Arthur and Birger, Chet and Thorvaldsd{\'o}ttir, Helga and Ghandi, Mahmoud and Mesirov, Jill P. and Tamayo, Pablo},
  title = {The {Molecular Signatures Database} {Hallmark} Gene Set Collection},
  journal = {Cell Syst.},
  volume = {1},
  number = {6},
  pages = {417--425},
  year = {2015}
}

@article{azgp1,
  author = {Brooks, James D. and others},
  title = {Loss of Expression of {AZGP1} Is Associated With Worse Clinical Outcomes in a Multi-Institutional Radical Prostatectomy Cohort},
  journal = {Prostate},
  volume = {76},
  number = {15},
  pages = {1409--1419},
  year = {2016}
}

@article{msmb,
  author = {Dahlman, Anna and others},
  title = {Evaluation of the prognostic significance of {MSMB} and {CRISP3} in prostate cancer using automated image analysis},
  journal = {Mod. Pathol.},
  volume = {24},
  number = {5},
  pages = {708--719},
  year = {2011}
}

@article{virchow,
  author = {Vorontsov, Eugene and others},
  title = {A foundation model for clinical-grade computational pathology and rare cancers detection},
  journal = {Nat. Med.},
  volume = {30},
  number = {10},
  pages = {2924--2935},
  year = {2024}
}

@article{virchow2,
  author = {Zimmermann, Eric and others},
  title = {{Virchow2}: Scaling Self-Supervised Mixed Magnification Models in Pathology},
  journal = {arXiv:2408.00738},
  year = {2024}
}

@article{phikon,
  author = {Filiot, Alexandre and others},
  title = {Scaling Self-Supervised Learning for Histopathology with Masked Image Modeling},
  journal = {medRxiv},
  year = {2023}
}

@article{phikonv2,
  author = {Filiot, Alexandre and Jacob, Paul and Mac Kain, Alice and Saillard, Charlie},
  title = {{Phikon-v2}, A large and public feature extractor for biomarker prediction},
  journal = {arXiv:2409.09173},
  year = {2024}
}

@article{hibou,
  author = {Nechaev, Dmitry and Pchelnikov, Alexey and Ivanova, Ekaterina},
  title = {{Hibou}: A Family of Foundational Vision Transformers for Pathology},
  journal = {arXiv:2406.05074},
  year = {2024}
}

@inproceedings{mocov3,
  author = {Chen, Xinlei and Xie, Saining and He, Kaiming},
  title = {An Empirical Study of Training Self-Supervised Vision Transformers},
  booktitle = {Proc. ICCV},
  pages = {9620--9629},
  year = {2021}
}

@article{tcga,
  author = {Weinstein, John N and others},
  title = {The {Cancer Genome Atlas} {Pan-Cancer} analysis project},
  journal = {Nat. Genet.},
  volume = {45},
  number = {10},
  pages = {1113--1120},
  year = {2013}
}

@article{reactivestroma,
  author = {Ayala, Gustavo and others},
  title = {Reactive Stroma as a Predictor of Biochemical-Free Recurrence in Prostate Cancer},
  journal = {Clin. Cancer Res.},
  volume = {9},
  number = {13},
  pages = {4792--4801},
  year = {2003}
}

@article{fieldcancerization,
  author = {Nonn, Larisa and Ananthanarayanan, Vijayalakshmi and Gann, Peter H.},
  title = {Evidence for Field Cancerization of the Prostate},
  journal = {Prostate},
  volume = {69},
  number = {13},
  pages = {1470--1479},
  year = {2009}
}

\end{document}